\documentclass[showpacs,prl,twocolumn,aps]{revtex4}

\usepackage{amsmath}
\usepackage{amssymb}
\usepackage{latexsym}
\usepackage{amsfonts}
\usepackage{epsfig}
\usepackage{psfrag}

\newcommand{\braket}[2]{\left<#1|#2\right>}

\newcommand{\ul}{\underline}
\newcommand{\f}[1]{\mbox{\boldmath$#1$}}

\newcommand{\na}{\mbox{\boldmath$\nabla$}}
\newcommand{\bea}{\begin{eqnarray}}
\newcommand{\ea}{\end{eqnarray}}
\newcommand{\eea}{\end{eqnarray}}
\newcommand{\ord}{\,{\cal O}}
\newcommand{\sumint}[1]
{\begin{array}{c} \\
{{\textstyle\sum}\hspace{-1.1em}{\displaystyle\int}}\\
{\scriptstyle{#1}}
\end{array}}

\begin{document}

\title{Dynamically assisted Schwinger mechanism}  

\author{Ralf Sch\"utzhold$^{1,2}$, 
Holger Gies$^{3,4}$, and Gerald Dunne$^{5}$}

\affiliation{$^1$ Fachbereich Physik, Universit\"at Duisburg-Essen, 
D-47048 Duisburg, Germany
\\
$^2$ Institut f\"ur Theoretische Physik, Technische
  Universit\"at Dresden, D-01062 Dresden, Germany
  \\
  $^3$ Institut f\"ur Theoretische Physik, 
  Universit\"at Heidelberg, D-69120 Heidelberg, Germany
    \\
  $^4$ Theoretisch-Physikalisches Institut, 
  Universit\"at Jena, D-07743 Jena, Germany
    \\
  $^5$ Department of Physics, 
  University of Connecticut, Storrs Ct 06269, USA} 

\begin{abstract}
We study electron-positron pair creation {from} the Dirac vacuum 
induced by a strong and slowly varying electric field (Schwinger effect) 
which is superimposed by a weak and rapidly changing electromagnetic field 
(dynamical pair creation). 
In the sub-critical regime where both mechanisms separately are
strongly suppressed, their combined impact yields a pair creation rate
which is {dramatically} enhanced. 
Intuitively speaking, the strong electric field lowers the threshold for 
dynamical particle creation -- or, alternatively, the fast electromagnetic 
field generates additional seeds for the Schwinger mechanism.
These findings could be relevant for planned ultra-high intensity lasers.
\end{abstract}


\maketitle


As first realized by Dirac \cite{Dirac}, a consistent relativistic quantum
description of electrons necessarily involves negative energy levels, which
-- in the Dirac-sea picture -- are filled up in the vacuum state.
This entails the striking possibility of pulling an electron out of 
the vacuum by means of some external influence, such as a (classical) 
electromagnetic field \cite{Sauter}, where the remaining hole in the Dirac 
sea is then associated with a positron. 
Of course, to create such an electron-positron pair out of the vacuum,
one has to overcome the energy gap of $2mc^2$ between the filled and 
the empty levels.
There are basically two main mechanisms for doing so:
In a strong electric field $E$ over a sufficiently long distance $L$,
{``virtual''} electron-positron pair {fluctuations} may gain this energy
when $qEL\geq2mc^2$.
This pair creation process is called the Schwinger mechanism
\cite{Heisenberg+Euler,Schwinger} and can be understood as tunneling 
through the classically forbidden region (energy gap).
Thus it is suppressed exponentially 
$\ord(\exp\{-\pi {E_{\text{S}}}/E\})$ 
for weak fields $E$, where 
${E_{\text{S}}}=m^2c^3/(\hbar q)$ 
is the Schwinger critical field. 
For $E\simeq E_{\text{S}}$, the work done by separating
the electron-positron pair over a Compton wavelength is of the 
order of the energy gap $2mc^2$. 
Alternatively, a classical time-dependent electromagnetic field will 
also create electron-positron pairs in general (dynamical pair creation).
However, if the frequency $\omega$ of the external field is not 
large enough, $\hbar{\omega}<2mc^2$, these non-adiabatic corrections 
correspond to higher-order (i.e., multi-photon) processes and are also 
suppressed exponentially {$\exp\{-\ord(1/\omega)\}$} 
for small $\omega$ \cite{Brezin}. 
These pair-production processes are fundamental predictions of quantum
electrodynamics (QED), but only the multi-photon production process has 
so far been observed experimentally: 
the positron data taken at the SLAC E-144 experiment have convincingly 
been explained by $n$-photon production with $n\simeq 5$ 
\cite{Burke:1997ew}. 
However, a verification of the Schwinger mechanism has still remained 
an experimental challenge \cite{Ringwald}. 
Since the Schwinger mechanism is non-perturbative in the field, its 
discovery would help exploring the non-perturbative realm of quantum 
field theory in a controlled fashion.  
Here, we propose a new mechanism which can help to overcome the 
strong exponential suppression.
The basic idea is similar in spirit to ideas in the study of 
atomic physics in strong fields, where new experimental and 
theoretical results show that controlled engineering of special
electric field pulse shapes can enhance certain interesting 
physical processes, such as high-harmonic generation and 
above threshold ionization (for reviews see \cite{becker+gerber}).

Many previous theoretical studies of pair production
\cite{Bunkin,Nikishov,Brezin,Popov,Gies:2005bz} 
have been motivated by the seminal work of Keldysh \cite{Keldysh} on
atomic ionization in time-dependent electric fields; in particular the
crossover between the two main mechanisms of pair creation due to
strong constant electric fields and due to those with
spatial or temporal variations has been of interest.
It turns out that spatial variations tend to diminish the pair creation
rate \cite{Nikishov,Gies:2005bz} whereas a time-dependence typically
increases the effect \cite{Popov,WLI}.
However, a realistic experimental situation is usually far more complex 
and may involve various frequency and amplitude scales over a wide range. 
This motivates us to study electron-positron pair creation in the 
presence of a strong and slow electric field plus weak and fast 
electromagnetic wiggles. 
We assume that the slow electric field $E$ is strong but still far below 
the Schwinger limit ${E_{\text{S}}}$, and that the frequency of the 
weak electromagnetic wiggles is smaller than twice the electron mass. 
As explained before, the pair creation rate of each effect separately 
is strongly suppressed in this case.
As we shall demonstrate below, however, their combined impact may be 
much stronger, i.e., yield an enhanced pair creation rate. 
These findings could be experimentally relevant in view of the
next-generation light sources \cite{ELI} aiming at approaching the
Schwinger limit via high-harmonic focusing, which typically 
generates a high-frequency tail.

\begin{figure}[ht]
\includegraphics[height=3.5cm]{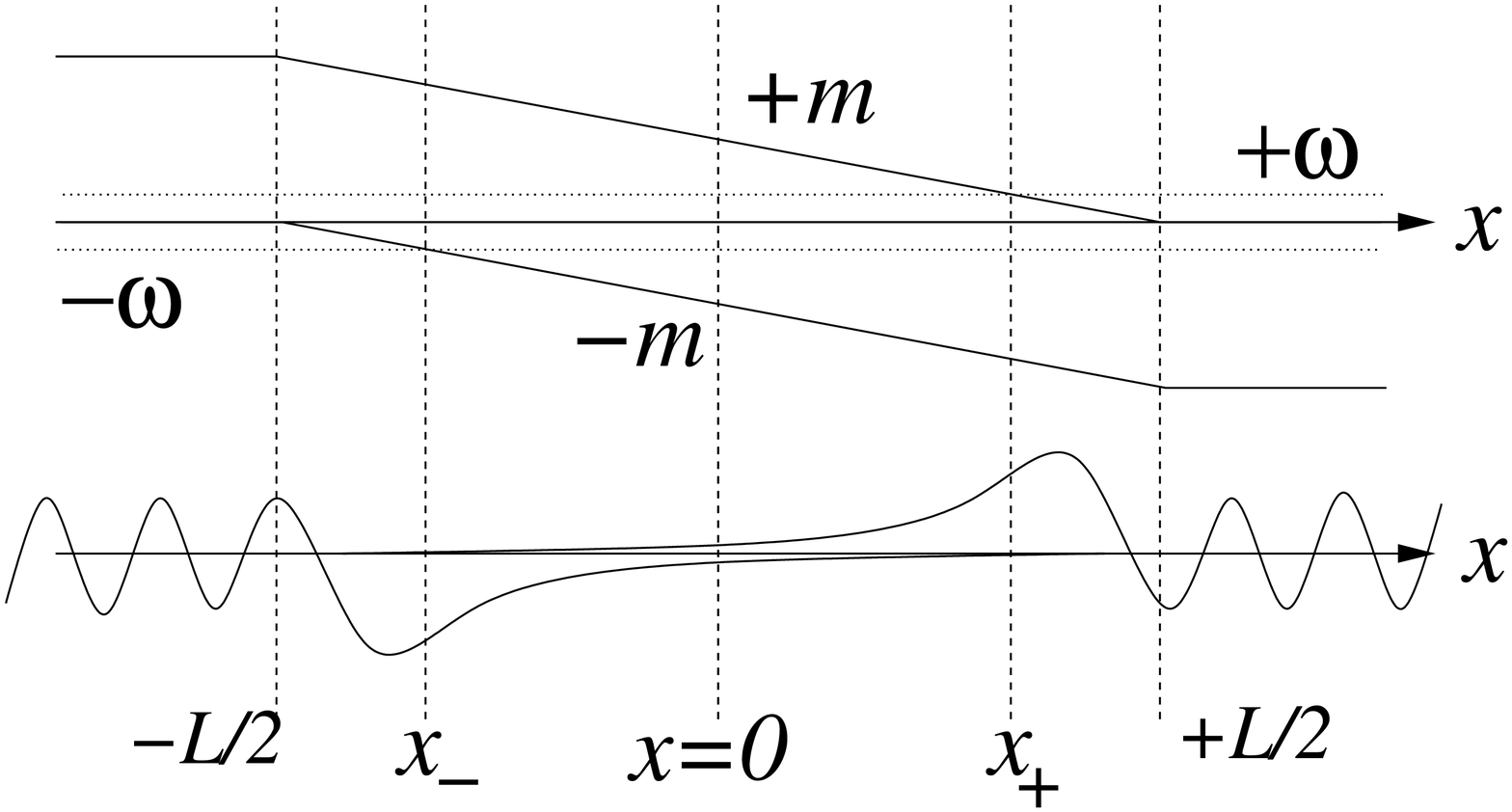}
\caption{\label{figure} Sketch (not to scale) of the level structure (top)
and the mode functions $u_I^\dagger$ and $v_J$ (bottom).
The upper and lower surface of the Dirac sea at $\pm m$ are denoted by solid 
lines, which are  distorted by the electric field $E$ in the interval 
$-L/2<x<+L/2$ with $qEL= 2m$ (top).
The horizontal dotted lines at $\pm\omega$ represent the electron/positron 
levels $u_I^\dagger$ and $v_J$ with the classical turning points at $x_\pm$. 
}
\end{figure}

For a first estimate, let us treat the strong and slow electric field
adiabatically and non-perturbatively (as our $\hat H_0$ problem) and the
weak and fast electromagnetic wiggles non-adiabatically and
perturbatively.
In terms of the field operator $\hat\Psi$ and the Dirac 
matrices $\f{\ul\alpha}$ and $\ul\beta$, the unperturbed Hamiltonian 
density reads ($\hbar=c=1$)
\bea
\label{H_0}
\hat{\cal H}_0=\hat\Psi^\dagger
\left(i\f{\ul\alpha}\cdot\na+m\ul\beta+V\right)
\hat\Psi
\,,
\ea
where $V(\f{r})=qA_0$ encodes the (approximately static) electric field.  
The perturbation Hamiltonian 
\bea
\label{H_1}
\hat{\cal H}_1=q\hat\Psi^\dagger
\f{\ul\alpha}\cdot\f{A}(t,\f{r})
\hat\Psi
\ea
contains the vector potential $\f{A}(t,\f{r})$ of the fast 
electromagnetic wiggles (scale separation).
The $\hat H_0$ problem can be diagonalized via the usual 
normal-mode expansion 
\bea
\label{normal-mode}
\hat\Psi(t,\f{r})=
\sumint{I} 
\hat a_I u_I(\f{r}) e^{-i\omega_It}+
\hat b_I^\dagger v_I(\f{r}) e^{+i\omega_It}
\,,
\ea
where the positive/negative energy spinor solutions $u_I(\f{r})$
and $v_I(\f{r})$ depend on spin $\sigma$ and wavenumber $\f{k}$,
which are combined {into} the index $I=\{\sigma,\f{k}\}$.
The electron/positron creation/annihilation operators 
$\hat a_I,\hat a_I^\dagger$ and $\hat b_I,\hat b_I^\dagger$
are time-independent. 
After insertion of (\ref{normal-mode}) into (\ref{H_1}), the 
pair-creation amplitude ${\mathfrak A}_{IJ}$ can be calculated 
in first-order perturbation theory 
\bea
\label{amplitude}
{\mathfrak A}_{IJ}
=
q \int d^4x\,
u_I^\dagger(\f{r}) 
\f{\ul\alpha}\cdot\f{A}(t,\f{r})
v_J(\f{r}) 
\,
e^{i\omega_It+i\omega_Jt}
\,. 
\ea
Within the present estimate, we are primarily interested in the size of
the exponent governing the exponential suppression of the above 
amplitude. 
Therefore, we study a 1+1 dimensional toy model, 
which should reproduce the exponent correctly, but disregard the
sub-leading pre-factor.
Let us consider a constant electric field $E$ over an interval of length 
$L$ with vanishing field outside, see Fig.~\ref{figure}.
For simplicity, we assume {$qEL\gtrsim 2m$}, i.e., we are just above the
threshold for the Schwinger effect.
In terms of the Klein paradox \cite{Klein} language, the case lies 
near the border between the intermediate and strong potential 
regime. 

For modes $I,J$ whose frequency sum $\omega_I+\omega_J$ corresponds 
to the typical frequency of the perturbation $\f{A}(t,\f{r})$, the 
exponential suppression of the pair-creation amplitude (\ref{amplitude})
is basically determined by the spatial overlap of the modes 
$u_I^\dagger(\f{r})$ and $v_J(\f{r})$. 
Assuming $\omega_I=\omega_J=\omega$ for simplicity
(other distributions  only induce a shift in $x$ but lead to the same 
result as long as $\omega_I+\omega_J=2\omega$), the spinor 
$u_I(\f{r})$ describes electron modes which are incident from the right and
totally reflected (due to $\omega<m$) by the strong field $E$, 
whereas $v_J(\f{r})$ corresponds to positron modes which are incident from 
the left and also totally reflected.  
The classical turning points are given by $x_\pm=\pm(m-\omega)/(qE)$, 
see Fig.~\ref{figure}.
As the electric field $E$ is assumed to lie far below the Schwinger limit,
we may employ the WKB approximation and estimate the exponential 
suppression by the integral of the eikonal between
the classical turning points 
\bea
\int\limits_{x_-}^{x_+} dx\,\sqrt{m^2-(qEx-\omega)^2}=
\frac{m^2}{4qE}\times 
f\left(\frac{\omega}{m}\right)
\,,
\label{eq5}
\ea
with $f(\chi)=\pi+2\arcsin(1-2\chi)+4\sqrt{\chi}(1-2\chi)\sqrt{1-\chi}$,
which can be approximated by $f(\chi)\approx2\pi(1-\chi)$ in the relevant 
interval $\chi\in[0,1]$.
In the limit $\chi=0$, we exactly recover Schwinger's {exponential factor}  
$\exp\{-\pi {E_{\text{S}}}/E\}$ for the pair-creation 
probability in a static field.
For $\chi>0$, however, the exponent is reduced.
Intuitively speaking, the particles do not have to tunnel all the way 
from $-L/2$ to $+L/2$ because the frequency $\omega$ helps them to 
penetrate the strong field region up to $x_\pm$. 
In a dual picture, the particles tunnel through part of the gap, 
until they can be excited by frequency $\omega$. 
At $\chi=1/2$, we get exactly half the exponent, and hence the 
pair-creation rate is drastically enhanced 
$\exp\{-\pi{E_{\text{S}}}/(2E)\}$.
Note that, without the strong electric background field $E$, 
{a single photon} with $\omega=m/2$ would not have enough 
energy to create an electron-positron pair. 
Below threshold $\omega<m$, pair production requires multi-photon
  processes which occur at higher orders in the above expansion scheme.
For $\chi=1$, the exponent {in Eq.~(\ref{eq5})} vanishes as 
expected, since the electromagnetic wiggles have enough energy for 
pair creation $\omega=m$.

The above {approach based on the} scenario in Fig.~\ref{figure} has
the advantage of allowing arbitrary wiggles $\f{A}(t,\f{r})$, but 
has the drawback that multi-photon processes require high-order
calculations. 
Also realistic backgrounds $E(t,x)$ are difficult to handle even though 
the tunneling exponent is expected to be universal for slowly 
varying fields.
Inhomogeneous backgrounds as well as multi-photon physics can be dealt 
with by the worldline instanton technique \cite{WLI} which we apply 
to the following specific example: 
we consider a strong and slow electric field pulse superimposed by a 
weak and fast pulse, both spatially homogeneous,
\bea
\label{cosh}
\f{E}(t)=\frac{E}{\cosh^2(\Omega t)}\f{e}_z+
\frac{\varepsilon}{\cosh^2(\omega t)}\,\f{e}_z
\ea
with ${E_{\text{S}}}\gg E\gg\varepsilon>0$ and
$m\gg\omega\gg\Omega>0$.
With only one such pulse (say $\varepsilon=0$), the corresponding pair
creation rate can be computed analytically \cite{Nikishov}.  
For the superimposed dual-pulse form in (\ref{cosh}), we can compute 
the pair creation rate semi-classically using an analytic continuation 
to Euclidean time $x_4$ 
\bea
\label{tan}
A_3(x_4)=-i\frac{E}{\Omega}\,\tan(\Omega x_4)
-i\frac{\varepsilon}{\omega}\,\tan(\omega x_4)
\quad ,
\ea
with the tunneling exponent being related to the action of the
worldline instanton.
Starting with the worldline representation of the path integral, 
we may use the electron mass $m$ as a large parameter for the 
saddle-point approximation. 
The saddle points corresponding to the tunneling events are 
word-line instantons $x_\mu=[0,0,x_3(\lambda),x_4(\lambda)]$
which are closed loops in Euclidean space-time satisfying 
the equation of motion 
\bea
\left(\frac{dx_4}{d\lambda}\right)^2+
q^2
\left(\frac{E\tan(\Omega x_4)}{m\Omega}
+\frac{\varepsilon\tan(\omega x_4)}{m\omega}
\right)^2=1
\,,
\ea
where $d\lambda^2=dx_3^2+dx_4^2$ is the proper time. 
This equation describes the classical motion of a particle in a 
potential.
For small $\varepsilon$, the second term $\tan(\omega x_4)$ acts as an 
infinitely high rectangular well potential and just reflects instanton 
trajectories $x_4(\lambda)$ at the walls $\omega x_4=\pm\pi/2$.
Between the walls, we have an approximately harmonic oscillation due to 
$\Omega\ll\omega$ and thus $\tan(\Omega x_4)\approx\Omega x_4$. 
The structure of the solution $x_4(\lambda)$ depends on the combined 
Keldysh adiabaticity parameter
\bea
\label{kel}
\gamma=\frac{m\omega}{qE} 
\quad .
\ea
Note that the relevant Keldysh parameter in this multi-scale problem 
is formed out of the dominant frequency $\omega$ of the fast pulse on
the one hand, and the dominant field strength $E$ of the slow pulse on
the other hand.
For small $\gamma\ll 1$, we approach the pure Schwinger limit, 
whereas large $\gamma$ 
do \emph{not} correspond to a pure multi-photon regime \cite{Brezin} 
as measured in the SLAC E-144 experiment \cite{Burke:1997ew}; 
large $\gamma$ still involve both multi-photons of frequency $\omega$ 
as well as a non-perturbative dependence on $E$. 

For small Keldysh parameters $\gamma<\pi/2$, the instanton trajectories
do not reach the walls and reflection does not occur, i.e., the 
$\tan(\omega x_4)$-term has no impact.
In this case, the weak pulse is too slow to create pairs significantly 
and we essentially reproduce Schwinger's result. 
Beyond this threshold, $\gamma>\pi/2$, however, the instanton 
trajectories $x_4(\lambda)$ change due to reflection at the walls, 
and the instanton action becomes modified
\bea
{\cal A}_{\rm inst}=m\oint d\lambda
\left(\frac{dx_4}{d\lambda}\right)^2 
=4m\int\limits_0^{\lambda_*} 
\left(\frac{dx_4}{d\lambda}\right)^2 
\ea
due to the reflection points at 
\bea
\lambda_*=\frac{m}{qE}\,\arcsin\left(\frac{\pi}{2\gamma}\right)
\,.
\ea
Consequently, we obtain (for $\gamma\geq\pi/2$)
\bea
\label{threshold}
{\cal A}_{\rm inst} 
\,{\approx}\,
\frac{m^2}{qE}\,
\left[
2\arcsin\left(\frac{\pi}{2\gamma}\right)
+\frac{\pi}{2\gamma^2}\sqrt{4\gamma^2-\pi^2}
\right]
\,.
\ea
%

\begin{figure}[ht]
  \includegraphics[scale=0.5]{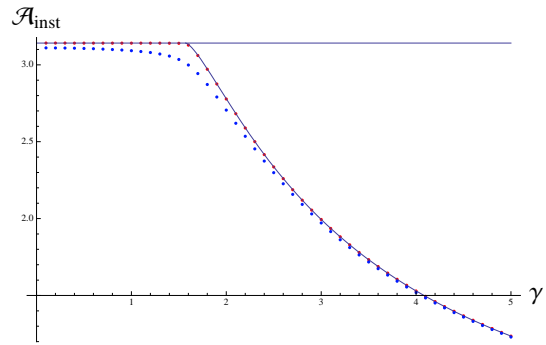}
  \caption{\label{figure2} Plots of the instanton action [in units of
    $m^2/(qE)$] for the electric field in (\ref{cosh}), computed using the
    wordline instanton method, and plotted as a function of the combined
    Keldysh parameter $\gamma$ defined in (\ref{kel}). The upper [red]
    dots correspond to $\omega=100\Omega$ and $E=100 \varepsilon$, while the
    lower [blue] dots correspond to $\omega=10\Omega$ and $E=10
    \varepsilon$. The solid lines show the Schwinger value of $\pi$,
    estimated in the text to be valid for $\gamma<\pi/2$, and the
    expression (\ref{threshold}), estimated in the text to be valid for
    $\gamma>\pi/2$. The numerical results agree very well with these
    estimates in the relevant limit where $E\gg \varepsilon$ and
    $\omega\gg\Omega$.  }
\end{figure}

At the threshold, $\gamma=\pi/2$, we reproduce the Schwinger value 
${\cal A}_{\rm inst}=\pi m^2/(qE)$, as one would expect.
Above the threshold, $\gamma>\pi/2$, the instanton action 
${\cal A}_{\rm inst}$ decreases significantly. 
For example, for $\gamma=\pi$, it is reduced by about 40\%. 
For $\gamma\to\infty$, it decays as $1/\gamma$ in agreement 
with the expected multi-photon behavior \cite{Brezin}. 
For larger $\varepsilon$, the threshold behavior becomes smoother, 
see Fig.~\ref{figure2}.
Since the pair creation probability, determined by the imaginary part 
of the effective action $\Gamma[A_\mu]=-i\ln\braket{\rm in}{\rm out}$, 
depends exponentially on the instanton action, i.e., the saddle-point 
value
\bea
\Im(\Gamma[A_\mu])\sim\exp\{-{\cal A}_{\rm inst}\}
\,,
\ea
such a reduction of ${\cal A}_{\rm inst}$ implies a drastic
enhancement of the pair creation probability
$\Im(\Gamma[A_\mu])$; 
e.g., a reduction of 40\% in the exponent could make the difference 
between a suppression of $10^{-10}$ and $10^{-6}$, which could 
mean a few electron-positron pairs per day, instead of one per year. 
Of course, one could also reduce ${\cal A}_{\rm inst}$ by a factor 
of two via doubling the field $E$.
However, such strong fields are at the edge of present experimental 
capabilities and focusing two ultra-high intensity pulses into the
same space-time region is much harder than superimposing the strong 
pulse and a weak high-frequency pulse.
Similarly, increasing the characteristic frequency of the ultra-high 
intensity pulse is much harder than just adding a weaker field of 
high frequency.
In fact, the envisioned generation mechanism (high-harmonic focusing) 
for the ultra-high intensity pulses typically induces a high-frequency
tail automatically (see below). 

Due to the rather general nature of the arguments used above, they are 
not restricted to the specific pulse in Eq.~(\ref{cosh}).
One would obtain analogous results for 
\bea
A_3(x_4)=-iEx_4f(\Omega x_4)
-i\varepsilon\,\frac{(\omega x_4)^{2m+1}}{[1-(\omega x_4)^{2n}]^l}
\,,
\ea
with arbitrary positive integers $l,m,n$ and suitable linear combinations 
of such fields. 
The electric field $E(t)$ is then an (even) polynomial in time 
over the numerator $[1+(\omega t)^{2n}]^l$; which might capture 
some more features of a laser pulse than Eq.~(\ref{cosh}). 
In this case, the threshold is at $\gamma=1$, but after rescaling
$\omega$, we again get Eq.~(\ref{threshold}).
This behavior is confirmed by a worldline instanton computation. 
In order to understand the general result, it might be helpful to 
consider the following picture:
The strong and slow pulse deforms the fermionic levels almost 
adiabatically.  
As a result, the expectation value of the {\em free} electron and 
positron number operator (valid for $A_\mu=0$) would scale with 
$E^2/m^4$ (plus adiabatic corrections $\dot E/m^2$ etc.) 
and thus give a rather large result.
However, this large number does not count {\em real} electrons/positrons 
(e.g., they do not annihilate) but mostly the instantaneous deformation 
of the ground state.
The number of {\em real} electron-positron pairs, left over after the 
pulse, scales exponentially {with} $\exp\{-\ord(1/E)\}$, 
i.e., non-perturbatively in $E$ \cite{hebenstreit:08}. 
On the other hand, the temporary deformation of levels during the strong
pulse can be exploited by the small wiggles, which can turn the deformed 
``virtual'' pairs into into real electron-positron pairs.  

The dramatic enhancement of the exponentially small pair-creation 
probability should be relevant for the present experimental efforts
aimed at the generation of light sources approaching the Schwinger 
limit, see, e.g., \cite{ELI}.
One of the main envisioned mechanisms for the final amplification 
stage is coherent high harmonic focusing: 
Let us imagine sending a laser pulse of ultra-high intensity onto a 
curved metal surface.
For a very high intensity, the Keldysh parameter of the laser is very 
small and hence electrons in the metal start to oscillate coherently
and ultra-relativistically. 
Thereby, they effectively form a relativistically flying mirror, 
which reflects the incident light with a large Doppler shift and 
thereby generates high harmonics up to large order $n$.
According to \cite{Pukhov}, the spectrum of these high harmonics 
is universal and the intensity of the $n$-th harmonic scales with 
$n^{-8/3}$ up to some cut-off, which is proportional to the third 
power of the Lorentz boost factor of the mirror and thus depends 
on the incident laser intensity.  

For example, an incident optical laser intensity of order 
$10^{22}$~W/cm$^2$ reachable in the near future would correspond 
to $n=\ord(10^5)$, which may range up to a significant fraction 
of the electron mass. 
Finally, the curvature of the metal surface allows us to focus the 
high harmonics into a small space-time region, such that the spatially 
and temporally compressed intensity might approach the Schwinger limit. 
In this scenario, the highest harmonics (near the cut-off) will not 
contribute to the peak intensity significantly -- but they still can 
enhance the pair creation probability drastically
(compared to what one would expect from the Schwinger mechanism alone).  

RS and HG acknowledge support by the DFG under grants SCHU~1557/1-2,3 
(Emmy-Noether program), Gi~328/4-1 (Heisenberg program), and the SFB-TR18. 
GD thanks the US DOE for support through grant DE-FG02-92ER40716.
 

\end{document}